\documentclass[12pt]{iopart}
\usepackage{iopams}  
\usepackage{graphicx,amssymb,amsfonts,latexsym,color,dcolumn,bm,epsfig,subfigure} 

\newcommand{\be}{\begin{equation}}
\newcommand{\ee}{\end{equation}}
\newcommand{\eenn}{\nonumber \end{equation}}
\newcommand{\beqn}{\begin{eqnarray}}
\newcommand{\eeqn}{\end{eqnarray}}

\def\bbm[#1]{\mbox{\boldmath $#1$}}

{\catcode`\|=\active\gdef\Braket#1{\left<\mathcode`\|"8000\let|\bravert {#1}\right>}} 
\def\bravert{\egroup\,\vrule\,\bgroup}

\begin{document}

\title[]{Bragg spectroscopy for measuring Casimir-Polder interactions with Bose-Einstein 
condensates above corrugated surfaces}
\author{Gustavo A. Moreno$^1$, Diego A. R. Dalvit$^2$, Esteban Calzetta$^1$}
\address{$^{1}$ CONICET and Departamento de F\'{i}sica, FCEN, Universidad de Buenos Aires, Ciudad
Universitaria, 1428 Buenos Aires, Argentina.\\$^2$ Theoretical Division, Los Alamos National Laboratory, Los Alamos, NM 87545, USA
}

\ead{morenog@df.uba.ar, dalvit@lanl.gov {\rm and} calzetta@df.uba.ar}

\begin{abstract}
We propose a method to probe dispersive atom-surface interactions by measuring via 
two-photon Bragg spectroscopy the dynamic structure factor of a Bose-Einstein 
condensate above corrugated surfaces. 
This method takes advantage of the condensate coherence to reveal
the spatial Fourier components of the lateral Casimir-Polder interaction energy.
\end{abstract}

\pacs{03.75.Kk, 03.75.Lm}

\submitto{\NJP}

\section{Introduction}
Dispersive atom-surface interactions are ubiquitous in several applications involving
cold atoms in proximity of bulk surfaces, including atom chips for quantum information processing, 
trapped neutral atoms and ions for precision measurements, and quantum reflection of ultracold matter from surfaces \cite{Review}. 
Such interactions arise from optical dipole forces
due to spatial gradients of the electromagnetic field caused by 
the reshaping of EM quantum vacuum fluctuations in the presence of material boundaries
\cite{Milonni}.
In recent years degenerate bosonic \cite{Antezza} and fermionic \cite{Inguscio2005} ultracold atomic gaseous 
systems have been proposed as ideal probes of dispersive atom-surface interactions due to their exquisite
control and characterization. In particular, frequency shifts of the center-of-mass of a 
Bose-Einstein condensate (BEC) have been used to measure equilibrium and non-equilibrium
Casimir-Polder forces \cite{Cornell}. Non trivial geometrical effects, such as the lateral Casimir-Polder force, 
could also be measured with a BEC in proximity to a corrugated surface
\cite{Dalvit}.

In this work we propose a method for probing atom-surface Casimir dispersive interactions based on the
modification of the excitation spectrum of a BEC brought close to a corrugated material surface. 
The quantum Casimir interaction induced by such a surface produces a periodic modulation of the trap potential
that qualitatively changes the condensate energy spectrum.  For example, a quasi one-dimensional condensate develops first order perturbation gaps in its
energy spectrum. The Bogoliubov states of the condensate which are significantly corrected have wavenumber commensurable with the Fourier components of
the Casimir potential, and thus the lateral Casimir-Polder force can be inferred from the modified spectrum. The Casimir-modified energy spectrum can be read out using two-photon Bragg spectroscopy techniques, which have been used to reveal the low energy spectrum of BECs trapped in elongated potentials \cite{Davidson} and optical lattices \cite{Inguscio2009}. In contrast to other proposals \cite{Dalvit} where the mechanical properties of the atomic cloud play an essential role in the description, this method relies on quantum properties of coherent matter such as the response of a many-body coherent interacting system to laser light. As we shall see, this effect is directly related to the low energy spectrum of the system and can be used to reveal lateral Casimir-Polder interactions of atoms with a surface. \\
We stress that we do not mean this contribution as a proposal for an experiment to be carried out in the immediate future. Our goal is to show how the distinctive features of a BEC, as opposed, e.g. to an incoherent gas, allow for new ways to explore atom-surface interactions. Indeed, we consider the novelty of transducing the Casimir lateral force into a band gap as the strongest point in this paper.\\
The paper is organized as follows. In Section II we review the problem of a single atom potential above a corrugated surface. We use this result in Section \ref{sec:spectrum} to determine the Casimir-Polder modified spectrum of an elongated BEC brought close to the surface. In Section \ref{sec:bragg} we discussed how the Casimir-modified energy spectrum
can be probed by two-photon Bragg spectroscopy. Finally we give numerical estimates of the effect in Section \ref{sec:numbers}.


\section{Casimir atom-surface interaction}

\begin{figure}[t]
\centerline{\epsfig{file=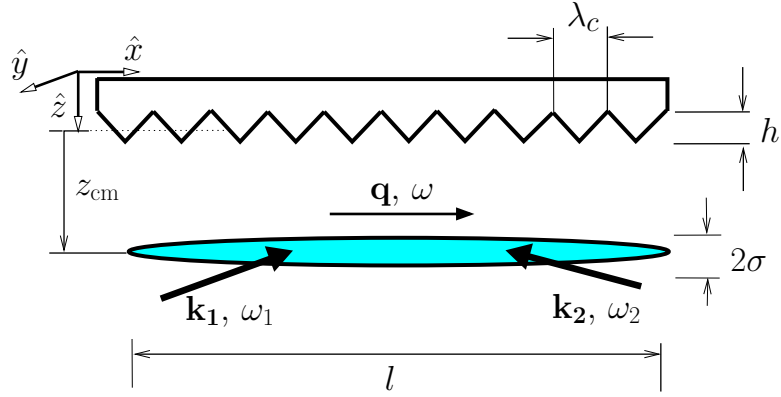,width=0.66\textwidth,angle=0}}
\caption{Set-up for probing Casimir atom-surface interactions by measuring the energy spectrum  via 
two-photon Bragg spectroscopy. The energy spectrum of an elongated 3D Bose-Einstein condensate trapped parallel 
to a corrugated surface is modified by the lateral component of the Casimir atom-surface interaction energy.
} 
\label{fig:idealized} 
\end{figure}

A ground-state atom at position ${\bf R}_A=(x_A,y_A,z_A)$ in front of a
corrugated surface (with surface profile $h(x,y)$ measured with respect to the plane $z=0$, see Fig. 1), 
is subjected to an atom-surface Casimir interaction energy $U$ due to the electromagnetic vacuum and thermal fluctuations that correlate the induced atomic electric dipole with fluctuating charges and currents in the surface (Fig. 1). 
For example, for a uni-axial corrugated surface with profile 
\begin{equation}\label{eq:profile_one}
h(x)=\sum_{j=1}^{\infty} h_j \cos(j k_c x), 
\end{equation}
where $h_j$ are the Fourier 
components of the profile and $\lambda_c=2\pi/k_c$ is the corrugation period, the interaction energy can be split as 
\begin{equation}
U(x,y,z)=U_N(z)+U_L(x,z), 
\end{equation}
where $U_N(z)$ leads to a {\it normal} force (for flat
surfaces it corresponds to the usual van der Waals/Casimir-Polder forces) \cite{Milonni} and $U_L(x,z)$ leads to 
a {\it lateral} force that appears only for non-planar surfaces \cite{Dalvit}. The first-order expansion of $U_L$ in powers of $h$ is 
\begin{equation}
U^{(1)}_L(x_A,z_A) = \sum_{j=1}^{\infty} h_j \cos(j k_c x_A) g(j k_c, z_A) ,
\label{interaction}
\end{equation}
where $g(k,z)$ is the response function \cite{Dalvit} containing information about the atomic response and about the geometry and optical response of the surface:
\begin{eqnarray}
g(k,z_A) &&= \frac{\hbar}{\epsilon_0c^2} \int_0^{\infty} \frac{\rmd\xi}{2\pi}
\xi^2\alpha(\rmi\xi)\int \frac{\rmd^2{\bf k}'}{(2\pi)^2} a_{{\bf k}',{\bf k}'-{\bf k}}
\label{g-def} \nonumber \\
a_{{\bf k}',{\bf k}''} &&= \frac{\exp[-(\kappa'+\kappa'') z_A]}{2\kappa''} \,\sum_{p',p''}
\hat{\bbm[\epsilon]}^{+}_{p'}({\bf k}') \cdot \hat{\bbm[\epsilon]}^{-}_{p''}({\bf k}'')R^{(1)}_{p'p''}({\bf k}', {\bf k}''),
\nonumber
\end{eqnarray}
where $\kappa=\sqrt{\xi^2/c^2+k^2}$. We remark that this expansion is valid only when $h(x)$ is the smaller length scale in the problem; for non-perturbative results see \cite{ContrerasReyes}. In the last expression
$\alpha(\rmi \xi)$ is the dynamic polarizability of the atom along imaginary frequencies, $\hat{\bbm[\epsilon]}^{\pm}_{p'}({\bf k}')$
are polarization vectors for incoming and reflected EM fields on the surface, and  $R^{(1)}_{p'p''}({\bf k}', {\bf k}'')$ are first-order
reflection matrices of EM fields impinging on the surface (see \cite{ref,Dalvit} for details). As discussed in \cite{Messina}, geometry
and conductivity corrections are approximately disentangled. The response function can be written as
\begin{equation}
g(k,z)=\rho(k,z) \eta_F(z)  F^{(0)}_{\rm CP}(z),
\end{equation}
where $\rho(k,z)\equiv g(k,z)/g(0,z)$ contains geometry corrections, and is an exponentially 
decaying function of the single variable ${\cal Z} =k z_A$ (for ${\cal Z}\gg 1$). Real material corrections are encapsulated in 
$\eta_F$ ($0\leq \eta_F \leq 1$), that is the conductivity correction to the normal component of the 
Casimir-Polder force  $F^{(0)}_{\rm CP}$ in the planar perfect reflector geometry. 
In the limit of separations much larger than the corrugation period ($k_c z_A \gg 1$), the exponential
decrease of $g$ implies that the $j=1$ term dominates in (\ref{interaction}), resulting in a effectively sinusoidal 
potential 
\beqn
U^{(1)}_{\rm L} = h_1 \cos(k_c x_A) g(k_c z_A)&&(k_c z_A \gg 1)\,\, .
\eeqn
Note that the same result holds for a height profile with a single Fourier component in a decomposition such as equation (\ref{eq:profile_one}), so for large enough separations ($z_A \gg \lambda_c$) different corrugation profiles become indistinguishable.

\begin{figure}[t]
\centerline{\epsfig{file=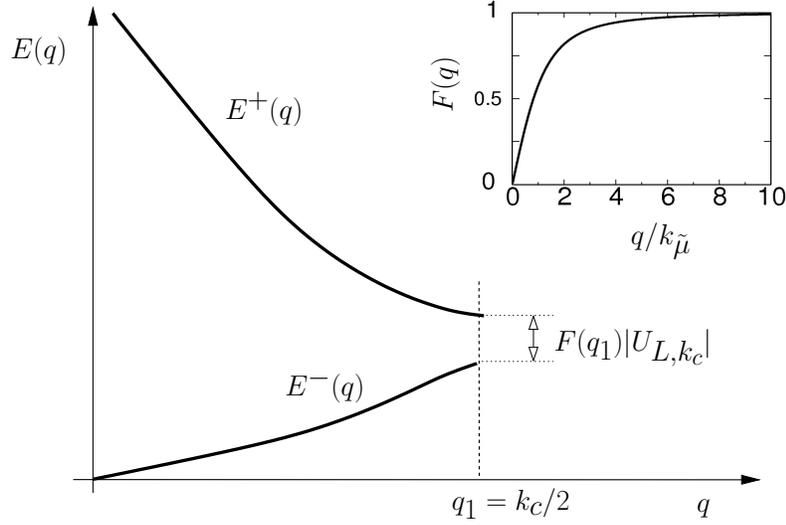,width=0.66\textwidth,angle=0}}
\caption{Modified energy spectrum of an elongated BEC trapped parallel to a surface
in the presence of a weak periodic lateral Casimir atom-surface interaction
The inset shows the function $F(q)$ that modulates the energy gaps $\Delta E_{q_n} = F(q_n) |U_{L,nk_c}|$ of the
unperturbed Bogoliubov spectrum.}
\label{fig:spectrum}
\end{figure}


\section{Casimir-modified BEC energy spectrum}\label{sec:spectrum}

Instead of considering the effect on a single atom we now compute the low energy spectrum of an interacting cloud of condensed atoms in presence of the corrugated surface. Consider a cigar-shaped BEC trapped by an axially-symmetric harmonic external potential, so that it is parallel to the corrugated surface, as shown in Fig.1.
The Casimir atom-surface interaction affects the mean-field dynamics of the condensate \cite{Antezza}, governed by the Gross-Pitaevskii equation (GPE)
\begin{eqnarray}
\rmi \hbar \partial_t \varphi &=& - (\hbar^2/2m) \nabla^2 \varphi + [U_N(z)+U_L(x,z)] \varphi \nonumber \\
&& + (m/2) (\omega_r^2 r^2+\omega_x^2x^2) \varphi+ g |\varphi|^2 \varphi,
\end{eqnarray}
where $\varphi$ is the condensate wavefunction, $m$ is the atomic mass, $g=4 \pi \hbar^2 a/m$, $a$ is the s-wave scattering length, and $\omega_r$ ($\omega_x$)
is the radial (axial) trapping frequency, $\omega_r \gg \omega_x$ \footnote[1]{Although Casimir forces are known to be non-additive, the fact that the condensate is a dilute object
justifies the computation of the total Casimir BEC-surface force as a sum over the Casimir forces between the surface and the individual atoms in the condensate.}. 
This interaction also modifies the structure of Bogoliubov fluctuations around the mean field solution and the corresponding energy
spectrum. Since the Casimir atom-surface interaction is a small perturbation to the external trapping potential, we will calculate the modifications to the BEC spectrum
in first order perturbation theory. 

In principle, one can start from the unperturbed Bogoliubov spectrum of the prolate elongated BEC, which has been calculated
numerically in \cite{Tozzo}. In the small wavelength limit $1/q \ll L$ (where $q$ is the axial quasi-particle momentum) the spectrum can be well described by the
discrete multibranch spectrum $E_{n,m}(q)$ of an infinitely long cylindrical condensate, where $n=0,1,2,\ldots$ is the radial quantum number and $m$ the radial
angular momentum \cite{Fedichev}.  Approximate analytical expressions for this spectrum can be found in some limiting cases which we analyse in the following two subsections.

\subsection{Low energy excitations and small chemical potential}

We first consider the case 
\begin{equation}
\mu - \hbar \omega_r \ll 8 \hbar \omega_r. 
\end{equation}
In this situation the radial confinement is so tight that the
dynamics of the BEC wavefunction is effectively 1D, the radial dynamics being ``frozen".  The Thomas-Fermi (TF) approximation for the radial dynamics is not valid in this regime, so we describe the effective dynamics writing the 3D wavefunction $\varphi$ in a basis $\{ f_n(r) \}$ of eigenfunctions
of the radial operator $-(\hbar^2/2 m) \Delta_r + m \omega_r^2 r^2/2$, namely 
\begin{equation}
\varphi = \sum_n f_n(r) \phi_n(x,t) ,
\end{equation}
(symmetry imposes no angular dependence). Projecting onto the fundamental
radial mode $f_0(r)$, it follows that the axial wavefunction $\phi_0(x,t)$ satisfies a 1D GPE
with an effective potential 
\begin{equation}
V_{\rm eff}(x)=\hbar \omega_r+ U_N(z_{\rm cm}) + U_L(x,z_{\rm cm}) ,
\end{equation}
and an effective interaction
$g_{\rm eff}=g/2 \pi \sigma^2$, with $\sigma^2=\hbar/m \omega_r$ (note we have approximated $z$ by the BEC center-of-mass 
position $z_{\rm cm}$; in a typical situation $U_N(z_{\rm cm})\ll \hbar \omega_r $). The non-linear coupling of $\phi_0$ to higher order modes $\phi_n$ can be neglected when $\mu - \hbar \omega_r \ll 8 \hbar \omega_r$, as can be seen when projecting the equation onto $f_0$.
When the typical axial length $l$ verifies $l \gg \lambda_c$, the condensate behaves locally as an interacting quasi 1D cold atomic gas in the presence of a weak 
Casimir atom-surface potential. The effect of the slowly varying axial 
external potential $m \omega_x^2 x^2/2$ will be incorporated in Section 4 using a local density approximation (LDA). In the absence of the surface, the energy spectrum is given by the Bogoliubov spectrum
for a quasi 1D homogeneous BEC, namely 
\begin{eqnarray}
E_{n,m}(q) &\approx &E_{n=0,m=0}(q) \approx  E_{\rm B} (q) \nonumber \\
&=& \sqrt{(\hbar^2 q^2/2 m) (\hbar^2 q^2/2 m + 2 \tilde \mu)},
\end{eqnarray} 
where $\tilde \mu=\mu-\hbar \omega_r-U_N(z_{\rm cm})$. 
To study the Casimir-modified spectrum we express the 1D BEC wavefunction as 
\begin{equation}
\phi(x,t)=\exp \left(-\rmi\frac{\mu t}{\hbar}\right) [\phi_{\rm TF}(x) + \delta\phi(x,t)], 
\end{equation}
where 
\be
\phi_{\rm TF}(x)=\{ [\tilde \mu - U_L(x,z_{\rm cm}) ] / g_{\rm eff} \}^{1/2}
\ee
is the TF mean field solution to the GPE above (valid when $\tilde \mu$ is greater than the typical kinetic energy due to spatial gradients), and
$\delta\phi(x,t)=u(x) \exp(-\rmi \frac{E t}{\hbar}) + v(x) \exp(\rmi \frac{E t}{\hbar})$ are the Casimir-modified
Bogoliubov excitations. These are solutions to 
\begin{eqnarray}
E u &=&-\frac{\hbar^2}{2m} \frac{d^2u}{dx^2} + (\tilde \mu - U_L(x,z_{\rm cm}))(u+v^*) , \nonumber \\
-E v &=& -\frac{\hbar^2}{2m} \frac{d^2v}{dx^2} + (\tilde \mu -  U_L(x,z_{\rm cm}))(u^*+v) . 
\end{eqnarray}

We now solve these equations to first order in powers of $U_L$. We write the Casimir-modified BEC energy spectrum as 
\begin{equation}
E(q)=E^{(0)}(q) + E^{(1)}(q) + \ldots. 
\end{equation}
Zeroth-order eigenfunctions are plane waves, namely
\begin{equation}
\begin{array}{ccc} u^{(0)}(x) &=&\sum_q u^{(0)}_q \,\exp(\rmi q x) \,\, ,\\ v^{(0)}(x)&=& \sum_q v^{(0)}_q\,\exp(\rmi q x) \,\, ,\end{array} 
\end{equation}
and the corresponding spectrum $E_q^{(0)}$
is equal to the Bogoliubov one, $E_{\rm B}(q)$. Expressing the Casimir energy $U_L(x,z_{\rm cm})$ in a cosine Fourier series 
(e.g., as in the small-$h$ limit, Eq.(\ref{interaction})), it follows that this weak periodic perturbation opens gaps in
the unperturbed energy spectrum at momenta $q_n=\pm n k_c/2$ ($n=0,1,\ldots$). As long as each Fourier component
$|U_{L,nk_c}| \ll E_q^{(0)}$ (which is consistent with the perturbative expansion), modes with different values of $n$ are effectively
uncoupled, and the gap for any fixed $n$ is obtained by solving the eigenvalue problem for degenerate unperturbed
states $n k_c/2$ and $-n k_c/2$. Solving the two-state problem for almost degenerate states $n k_c/2+\epsilon$ and
$-n k_c/2+\epsilon$, it is easy to find to first-order the energy branches (i.e. Bloch bands) $E^{\pm}(q)$ and the energy gaps between them on the border of the first Brillouin zone:
\begin{equation}\label{eq:gapCP}
\Delta E_{q_n}= |U_{L,nk_c}| \times F(q_n) \; ; \; F(q)=T_q/E^{(0)}_q ,
\end{equation}
where $T_q$ is the free kinetic energy
\begin{equation}\label{eq:freeKE}
T_q=\hbar^2 q^2/2m \,\, ,
\end{equation}
and $F(q)$ is a dimensionless suppression factor, plotted in Fig.2 together with the energy branches. Note that $F(q)\rightarrow 1$ for $q \gg k_{\tilde \mu}=(2m \tilde \mu/\hbar^2)^{1/2}$, corresponding to the particle-like region of the spectrum. For $q/k_{\tilde\mu} \ll 1$,
$F(q) \rightarrow 0$ \cite{Sorensen}.
Note that $x$-independent terms in the Casimir energy (like $U_N$) do not affect the energy gaps, and therefore
cannot be probed by Bragg spectroscopy. 
Our result (16) for the energy gaps due to the Casimir-Polder interaction is equivalent to those derived in previous studies of BECs in periodic potentials \cite{Sorensen}.

\subsection{Low energy excitations and large chemical potential}

For systems with higher densities the typical situation becomes $\mu \gg \hbar \omega_r$. In this case the radial dynamics can be described via the TF approximation, and
the unperturbed spectrum can be expanded in powers of $q R$ (with $R=(2 \mu/m\omega_r^2)^{1/2}$ the radial TF radius) \cite{Zaremba}
\begin{equation}
E^2_{n,m=0}(q)=2 (\hbar \omega_r)^2 n (n+1) + (q R)^2 \left(\frac{\hbar\omega_r}{2}\right)^2 + O((q R)^4).
\end{equation}
The lowest mode ($n=0$) corresponds
to axially propagating phonons with a speed of sound smaller by a factor $\sqrt{2}$ than the Bogoliubov speed of sound in the case a). 
Proceeding as before, one can compute the first-order energy gaps produced by the Casimir atom-surface interaction acting on the radially confined BEC in the high density limit, which results in
$\Delta E_{q_n} = (3 \hbar \omega_r / 4 \mu) \times (k_c R/2) \times |U_{L,nk_c}| $. Therefore, for large chemical potentials  the Casimir-induced energy gaps
are so small that they cannot be detected via Bragg spectroscopy (see below). It is thus convenient to consider condensates with relatively
small particle densities. 

\subsection{Beyond periodic corrugations}

So far we have considered the simplest case of a uniaxial corrugated surface, with Fourier spectrum $H(k_x) \propto \delta(k_x-k_c)$.
A similar procedure to the one described above can be followed for surfaces with more general corrugation profiles.
For example, if the surface may be described by two fundamental wavenumbers $k_{c1}$ and $k_{c2}$, namely
\begin{equation}\label{eq:profile}
h(x)=\sum_j h^{(1)}_j \cos(j k_{c1} x) \,+\sum_j h^{(2)}_j \cos(j k_{c2} x) , 
\end{equation}
one can apply the same calculation as in the single uniaxial case
provided the two-state problem defined for each fundamental wavenumbers $k_{c1}$ and $k_{c2}$ result independent of each other. This approach fails when the wavenumbers are close enough, because then the two sets of states will mix through the Casimir-Polder interaction. Thus, there will be a minimum separation in momentum space, say $\delta k_{\rm min}$, such that if $k_{c1}$ and $k_{c2}$ satisfy $|k_{c1}-k_{c2}| \gg \delta k_{\rm min}$ the two $2 \times 2$ problems are independent, but when this condition is not satisfied the first order energy correction will have to be computed taking into account that one is no longer dealing with two uncoupled systems. It is not difficult to see that the latter case yields to a $6 \times 6 $ problem; however,  in such cases, the relation between the energy gaps and the spatial Fourier components of the potentials may not be invertible. 

In fact, for certain surfaces, 
notably for those with stochastic roughness, this uncoupling condition can easily break down, and the method
proposed in this paper does not work. One can estimate the width $\delta k_{\rm min}$ on dimensional grounds. Taking into account that the perturbative parameter is $U_L/E^{(0)}$,  the minimum width should be of order $\delta k_{\rm min} \approx k_{c1,2} (U_L/E^{(0)})_{k_{c1,2}}$. We have verified that this is a good estimate by a direct diagonalization of the exact problem. Note also that this gives the minimum difference in momentum space that can be resolved when two fundamental wavenumbers $k_{c1}$ and $k_{c2}$ are present, and is the reason why Bragg spectroscopy of the low-energy BEC spectrum cannot resolve the Fourier components of the Casimir potential $U_L(k_x,k_y)$ produced by a rough surface. In the following, we will restrict ourselves to the simplest uniaxial corrugated case.

\begin{figure}[t]
\centerline{\epsfig{file=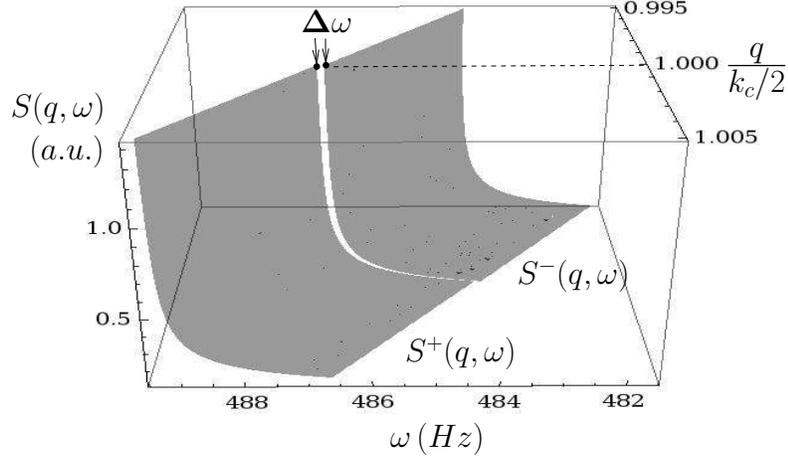,width=0.66\textwidth,angle=0}}
\caption{Dynamic structure factor (in arbitrary units) as a function of the detuning $\omega$ and the wave vector $q/(k_c/2)$. The parameters are chosen for $^{87}$Rb such that $\tilde \mu = \hbar\, 3.1$ kHz, $U^{(1)}_{L,k_c} \, f(k_c/2) = \hbar\, 0.11$Hz,  and $\lambda_c=2 \pi/k_c=9.75 \mu$m (see text).
}
\label{fig:spectrum}
\end{figure}


\section{Bragg spectroscopy of the Casimir potential}
\label{sec:bragg}

Consider two probe laser fields of frequency $\omega_1$ and $\omega_2$ and linear 
momentum ${\bf k}_1$ and ${\bf k}_2$ in the Bragg configuration of Fig. 1,
\begin{equation}
{\bf q}=q {\hat{\bf x}}= {\bf k}_1-{\bf k}_2 \; ; \; \omega=\omega_1-\omega_2.
\end{equation}
Bragg spectroscopy is an ideal tool for probing the BEC
energy spectrum via the measurement of the
dynamic structure factor (DSF) at zero temperature. The homogeneous DSF is found to be \cite{pitstri}
\begin{equation}\label{eq:DSFhomog}
S(q,\omega)=\frac{N \hbar^2q^2}{2 m E_{\rm B}(q)} \;  \delta(\hbar \omega - E_{\rm B}(q)), 
\end{equation}
where $N$ is the total number of BEC atoms. A similar expression is found for the Casimir-modified energy spectrum (calculated above neglecting the effect of the axial trapping potential $\propto \omega_x^2 x^2$), and furthermore the effect of the axial trapping potential can be incorporated via LDA averaging over the TF axial density profile \cite{pitstri}. The average can be performed via the integration of the DSF for the Casimir-modified spectrum using the local density profile given by $[1- (2x/l)^2]$. Performing the integral one finds two branches for the DSF, denoted below by $S^\pm(q,\omega)$,
\begin{equation} \label{eq:DSFprop}
S^\pm(q,\omega) \propto  \left[ \frac{\partial E^\pm(x,q)}{\partial x} \Big{|}_{x^*} \right]^{-1},
\end{equation}
where each branch $S^\pm(q,\omega)$ is associated with one energy branch through the relation $\hbar \omega = E^\pm(x^*,q)$. This last 
equation determines implicitly $x^*=x^*(\omega)$, that is to be used in equation (\ref{eq:DSFprop}) together with the local spectrum, which is defined by 
\begin{eqnarray}
E^\pm(x,q) &=& E^{(0)}(x,q) \pm \frac{T_q }{2 E^{(0)}(x,q)} \;  U_{L, n k_c}, \nonumber \\
E^{(0)}(x,q) &=& \sqrt{T_q^2+2 T_q \tilde \mu \left[ 1-  \left( \frac{2x}{L} \right)^2 \right] }, 
\end{eqnarray}
where $T_q$ is given by (\ref{eq:freeKE}). This gives a function which is not proportional to a delta-function 
but is still divergent when $\hbar \omega = E^\pm(0,q)$, which means there is a resonance when $\hbar \omega$ is the local energy \emph{at the origin}. 
Fig. 3 shows the two branches of the dynamic structure factor resulting from the Casimir atom-surface interaction for particular values of the parameters.

The actual observable in Bragg spectroscopy is not $S(q,\omega)$ but  the total momentum $P_X$ transferred to the BEC, whose equation of motion 
is given by \cite{Blakie}
\begin{eqnarray}
 &&\frac{\rmd P_X}{\rmd t}=-m\omega _{x}^{2} X + \sum_{n,i}  U_{L,n k_c} (n k_c) \langle \sin(n k_c x_i) \rangle + \nonumber \\ &&+\frac{\hbar q V_{\rm B}^2}{2}
\int \rmd \omega '\, \left[ S(q,\omega') - S(-q,-\omega ')\right] \,\frac{\sin ([\omega -\omega ']t)}{\omega -\omega '}.
\end{eqnarray}
The first term is due to the trapping potential ($X=\sum_i x_{i}$, with
$x_i$ the $x$-coordinates of the individual atoms in the BEC), 
the second term is the Casimir lateral energy $U_L$, 
and the last term is the response to the Bragg lasers which are assumed to have a Heaviside-theta envelope ($V_{\rm B}$ is the amplitude of the electric field of the superposed Bragg lasers). The time derivative of $P_X(t)$ is proportional to $S(q,\omega)$ for long enough pulses, that is of duration $\tau$ larger than the inverse of the typical energy scale $E=E_{\rm B}(k_c/2)$, provided that $\hbar \omega_x \ll E$
and $U_L$ is negligible. This last condition is verified
within our perturbative expansion, $U_{L,nk_c} \ll E$, which is further improved by partial cancellation of the sine terms in the second term above.
Thus, in the case we are considering, the resonances of the dynamic structure factor at fixed $q$ reveal the Casimir-modified energy spectrum. This gives an indirect measure of the Casimir-Polder interaction because once the gap in the spectrum ($\Delta \omega$) has been measured, the Fourier coefficient of the CP potential can be found using the relation (\ref{eq:gapCP}). Note that this fact does not depend on the approximations we have done to find an explicit expression for the CP lateral potential to first order in the corrugation amplitude.


\section{Numerical estimations and Discussion}
\label{sec:numbers}
Even if we do not mean this contribution as a blueprint for an experiment to be carried out with present technology, it is relevant to discuss whether such an experiment would be feasible in principle. In this Section we shall provide estimates for the strength of the effect in a typical experimental configuration. We shall not go into matters of experimental technique, such as how to sustain an adequate alignment of the elongated trap potential with respect to the surface, but rather focus on what can be said about the achievable band gap based on fundamental physics. 

On one hand, it is convenient to open gaps at large values of $q$ (short wavelength modes) in order
to maximize $F(q)$. However, on the other hand, this would imply exponentially suppressed Fourier
components of the Casimir-Polder lateral potential $|U_{L, q}|$. Therefore, the optimal parameters
will result from a compromise between the two factors in Eq.(\ref{eq:gapCP}). 
Let us evaluate the Casimir atom-surface lateral potential and the corresponding energy gaps in the
BEC spectrum for a benchmark configuration. Consider a sinusoidal uni-axial corrugated surface with corrugation wavelength 
$\lambda_c=2 \pi/k_c=9.75 \mu$m and corrugation amplitude $h=1\mu$m. 
In the following we will assume that the surface is separated by $z_{\rm cm}=3 \mu$m from a cigar-shaped
$^{87}$Rb condensate with $N=10^4$ atoms, trapped in an axially symmetric potential with trapping frequencies
$\omega_x=  2 \pi \times 0.83$ Hz and $\omega_r = 2 \pi \times 2.7$kHz. For this trapping frequency the radius of the BEC is $\sigma = 0.2 \mu$m. The chemical potential $\tilde \mu$ and the TF axial length $l$ are determined by the relations
\begin{equation}
N = \int \phi_{0, {\rm TF}}^2(x) \rmd x \; ; \;
\tilde \mu = \frac{1}{2}m \omega_x^2(l/2)^2. 
\end{equation}
That is,  $l/2=(3 g_{\rm eff}N / 2 m \omega_x^2)^{1/3}=408\mu$m 
and $\tilde \mu = (m\omega_x^2/8)^{1/3} (3 g_{\rm eff} N/2)^{2/3}=2 \pi \hbar  \times  493$Hz. For these parameters, $\tilde \mu \ll 8 \hbar \omega_r$, so that
we are under the conditions of subsection A of section III. The Thomas-Fermi approximation in the axial
direction is also satisfied because the relevant kinetic energy is $T_{q_1=k_c/2}=2 \pi \hbar \times 6.05$Hz $\ll \tilde \mu$. The typical Bogoliubov energy is  $E_{\rm B}(q_1)=2 \pi \hbar \times 77$Hz, 
and the suppression factor is $F(q_1)=0.08$. 

In order to compute the order of magnitude of the dispersive atom-surface energy, we first
consider the ideal case of a perfectly reflecting corrugated surface ($\eta_F=1$). The Casimir potential is computed from 
Eq.(\ref{interaction}) (note the caveat that,  for the chosen geometrical parameters $h/z_{\rm cm} \approx 0.33$, we are at the border of the validity of the first order approximation; the exact, non-perturbative expression can be found in \cite{ContrerasReyes}).
In the retarded Casimir-Polder limit, $z \gg \lambda_A$ (where $\lambda_A$ is the typical atomic transition wavelength),
the atom lateral CP potential for the perfectly reflecting surface is given to first order in $h$ by $U_L^{(1)}(x, z)=h \cos(k_c x) g^{\rm perf}_{\rm CP}(k_c, z)$, 
where \cite{Dalvit}
\begin{equation}
g^{\rm perf}_{\rm CP}(k, z)=- \frac{3 \hbar c \alpha(0)}{8 \pi^2 \epsilon_0 z^5} \rme^{-{\cal Z}}
(1+{\cal Z} + 16 {\cal Z}^2/45 + {\cal Z}^3/45) ,
\end{equation}
with ${\cal Z}=k_c z$ and $\alpha(0)/\epsilon_0=47.3 \times 10^{-30} {\rm m}^3$ is the static polarizability of Rb atoms. 
Therefore, the Fourier coefficient $U^{(1)}_{L,k_c}=h g^{\rm perf}_{\rm CP}(k_c, z_{\rm cm})$ is approximately $2 \pi \hbar \times 0.22$Hz.
Corrections due to real material properties can be calculated from \cite{Messina}. For the atom-surface separations considered, 
the geometry correction factor $\rho$ is well approximated by the perfect reflector case (Fig. 3 of \cite{Messina}), and the reduction factor is $\eta_F \approx 0.9$ for gold and $\eta_F \approx 0.7$ for silicon (Fig. 4 of \cite{Messina}). Therefore
$U^{(1)}_{L,k_c}$ is approximately  $2 \pi \hbar \times 0.2$Hz and $2 \pi \hbar \times 0.16$Hz for gold and silicon surfaces, respectively.

So far we have dealt with the case of zero temperature Casimir atom-surface interactions. Thermal corrections to these interactions can be easily computed replacing the integral over frequencies by a sum over thermal Matsubara frequencies. 
Thermal effects start to be important for distances $z_A$ larger than the thermal wavelength of the photon, $\lambda_T= \hbar c / k_{\rm B} T$, where $T$ is the temperature of the environment, $T=T_E$ (assumed to be in thermal equilibrium with the surface at temperature $T_S=T_E$; see \cite{Antezzatemperature}). 
Other thermal effects may affect the
coherence length of the BEC in the 1D configuration,  yielding an upper bound on the temperature of the thermal cloud around the condensate, $T_{\rm BEC}$. 
Let us note that the surface and environment temperatures $T_S$ and $T_E$ are very different from the BEC temperature (typically hundreds of K against tenths of $n$K) and play completely different roles.  
In the quasi-1D regime considered here it can be shown \cite{pitstri} that the typical decay length of the coherence is given by $2n_1\hbar^2/k_{\rm B} T_{\rm BEC} m$, where $n_1$ is the one-dimensional density. Using the above parameters one finds that the temperature of the BEC should be on the order of the $n$K to preserve the axial coherence up to scales on the order of the size of the sample.     
However, we note that a finite phase coherence length (shorter than the axial size but larger than
the corrugation period) is sufficient to probe lateral Casimir-Polder forces. Thus, such extremely cold
BEC temperatures for maintaining global axial coherence may not be required.

Using the above estimations for the suppression factor and for the Casimir atom-surface energy, the gap in the energy spectrum $U^{(1)}_{L,k_c} \, F(k_c/2)$ is of the order of  $2 \pi \hbar \times 0.016$ Hz, both for ideal and real surfaces. 
This shows that in order to measure the lateral Casimir potential it is required to resolve a $2 \pi \hbar \times  0.016$Hz gap in a spectrum centered at $2 \pi \hbar \times 77$Hz. 
This could be achieved by Bragg spectroscopy if $\omega_x$,
which limits the maximum resolution, is low enough \footnote[2] {We assume that $\tau \omega_x < 1$ in order for LDA to be valid along the axial direction, and to avoid
possible sloshing of the BEC.}. For the typical value chosen before ($\omega_x=2\pi \times 0.83$Hz) the spectral resolution should reveal the sub-Hz structure. 
However, it should be noted that such high sensitivities have not been experimentally achieved yet \footnote[3]{A much larger signal can be attained when the BEC is placed closer to the surface.  Scaling the parameters given before to $z_{\rm cm}=0.7\mu$m, 
$\lambda_c=4\mu$m, and $h=50$nm, results in a gap of $2 \pi \hbar \times 3.98$Hz centered at $E=2 \pi \hbar \times 191$Hz. Although this energy range has been experimentally demonstrated
\cite{Davidson}, the minimum distance of a BEC to the surface at present is limited to $2\mu$m.}. Future improvements in cold atom technology
could bring within reach the detection of nontrivial geometry effects of quantum vacuum via Bragg spectroscopy of a Bose-Einstein condensate.

Let us now compare our proposed setup for measuring lateral Casimir interactions via Bragg spectroscopy with the method
of frequency shifts of the center-of-mass oscillations of the BEC, which was demonstrated in a measurement of the {\it normal} Casimir-Polder force \cite{Cornell} and proposed as a suitable method for the detection of the {\it lateral} Casimir-Polder interaction  \cite{Dalvit}.
The frequency shift method applied to measuring lateral forces has a limited spatial resolution due to the TF radii of the condensate.
Furthermore, if tighter configurations are considered in such a context, the relative frequency shift becomes smaller than the experimental resolution reported in \cite{Cornell}. For example, using the parameters proposed above one finds that the maximum relative change in 
the {\sl lateral}  frequency shift is about $7 \times 10^{-7}$, while the reported experimental sensitivity of those experiments was $5 \times10^{-5}$ \cite{Cornell}. In contrast, the tighter Gaussian configuration proposed here would give an improved resolution in the distance to the surface, and the axial spatial resolution would only be limited by the accuracy in determining the laser wavenumber differences and 
depends neither on the radial density profile nor in any oscillation amplitude. However, as pointed out before, both techniques for measuring lateral Casimir-Polder forces remain at present at the edge of detectability. 


\section{Concluding Remarks}

Geometry effects of the quantum vacuum, such as the lateral Casimir-Polder atom-surface interaction, modify the
energy spectrum of a BEC in close proximity to a corrugated surface. The qualitative differences in the lowest energy (phonon-like) band were characterized in this context and a possible experimental set up for measuring the effect was discussed. As we have shown, using Bragg spectroscopy to measure this effect seems challenging with present day technology but could become feasible in the near future, opening a new window on the physics of the interaction between surfaces and coherent matter.


\ack
We are grateful to M. Modugno for useful comments and to V. Bagnato, F. Dalfovo, P.A. Maia Neto and J. Schmiedmayer for discussions. GAM and EC are supported in part by CONICET, ANPCYT and UBA (Argentina). DARD is grateful to the support of the U.S. Department of
Energy through the LANL/LDRD Program for this work.

\section*{References}

\end{document}